\documentclass
[aps,pre,10pt,a4paper,superscriptaddress,twocolumn,lengthcheck,showpacs,nobibnotes,nofootinbib,floats]{revtex4}%
\usepackage{amsfonts}
\usepackage{amsmath}
\usepackage{amssymb}
\usepackage{graphicx}%
\ifx\pdfoutput\relax\let\pdfoutput=\undefined\fi
\newcount\msipdfoutput
\ifx\pdfoutput\undefined\else
\ifcase\pdfoutput\else
\ifx\paperwidth\undefined\else
\ifdim\paperheight=0pt\relax\else\pdfpageheight\paperheight\fi
\ifdim\paperwidth=0pt\relax\else\pdfpagewidth\paperwidth\fi
\fi\fi\fi
\begin{document}
\preprint{ }
\title{Spatiotemporal chaos and the dynamics of coupled Langmuir and ion-acoustic
waves in plasmas}
\author{S. Banerjee}
\email{santo.banerjee@polito.it}
\affiliation{Department of Mathematics, Politecnico di Torino, Turin, Italy.}
\affiliation{Micro and Nanotechnology Division, Techfab s.r.l., Chivasso, Italy.}
\author{A. P. Misra }
\email{apmisra@visva-bharati.ac.in}
\affiliation{Department of \ Physics, Ume\aa \ University, SE-901 87 Ume\aa , Sweden.}
\altaffiliation{Permanent address: Department of Mathematics, Visva-Bharati University,
Santiniketan-731 235, India.}

\author{P. K. Shukla}
\email{ps@tp4.rub.de.}
\affiliation{Institut f\"{u}r Theoretische Physik IV, Ruhr-Universit\"{a}t Bochum, D-44780
Bochum, Germany.}
\author{L.Rondoni}
\affiliation{Department of Mathematics, Politecnico di Torino, Turin, Italy.}

\begin{abstract}
A simulation study is performed to investigate the dynamics of coupled
Langmuir waves (LWs) and ion-acoustic waves (IAWs) in an unmagnetized plasma.
The effects of dispersion due to charge separation and the density
nonlinearity associated with the IAWs, are considered to modify the properties
of Langmuir solitons, as well as to model the dynamics of relatively large
amplitude wave envelopes. \ It is found that the Langmuir wave electric field,
indeed, increases by the effect of \ ion-wave nonlinearity (IWN). Use of a
low-dimensional model, based on three Fourier modes shows that a transition to
temporal chaos is possible, when the length scale of the linearly excited
modes is larger than that of the most unstable ones. The chaotic behaviors of
the unstable modes are identified by the analysis of Lyapunov exponent
spectra. The space-time evolution of the coupled LWs and IAWs shows that the
IWN can cause the excitation of many unstable harmonic modes, and can lead to
strong IAW emission. This occurs when the initial wave field is relatively
large or the length scale of IAWs is larger than the soliton characteristic
size. Numerical simulation also reveals that many solitary patterns can be
excited and generated through the modulational instability (MI) of unstable
harmonic modes. \ As time goes on, these solitons are seen to appear in the
spatially partial coherence (SPC) state due to the free ion-acoustic radiation
as well as in the state of spatiotemporal chaos (STC) due to collision and
fusion in the stochastic motion. The latter results the redistribution of
initial wave energy into a few modes with small length scales, which may lead
to the onset of Langmuir turbulence in laboratory as well as space plasmas.

\end{abstract}
\received{11 January, 2010}
\accepted{13 April, 2010}
\startpage{1}
\endpage{102}
\maketitle

\section{Introduction}
The formation of envelope solitons, through the nonlinear interaction of
high-frequency (hf) electric fields and low-frequency (lf) ion-acoustic waves
(IAWs) is one of the most interesting features, and is extensively studied
problem in the context of turbulence in modern plasma physics.\ Formation of
these structures not only plays a role in turbulence, but also for plasma
heating, particle transport etc., and so their properties are of fundamental
interest. In plasmas, such envelope solitons are well-known Langmuir solitons
that are hf Langmuir waves (LWs) trapped by the density troughs associated
with lf IAWs (see e.g., Refs.
\cite{Zakharov,Karpman1,Karpman2,Deeskow,Schamel1,Schamel2}). \

During the past few years several attempts have been made to investigate the
dynamics of solitons including dressed solitons \cite{Karpman2}, soliton
collapse \cite{Hadzievski}, nucleation of cavitons \cite{Doolen}, radiation of
Langmuir solitons \cite{Karpman1}, Landau damping in partially incoherent LWs
\cite{Fedele} etc. An experimental observation for the excitation of IAWs has
also been reported in a Langmuir turbulence regime by Prado \textit{et al}
\cite{Prado}.\ Moreover, an increasing interest has also been found to study
the Langmuir turbulence as a result of chaos
\cite{Misra1,Misra2,Misra3,Chain,Batra,Rizzato}. \ \ When the electric field
intensity is strong enough to reach the modified decay instability threshold,
the interaction between LWs and IAWs results to `weak turbulence', and then
the LWs are scattered off IAWs under $T_{e}>>T_{i},$ where $T_{(e,i)}$ is the
electron (ion) temperature. On the other hand, when the field intensity is so
strong that the modulational instability (MI) threshold is exceeded, the LWs
are then essentially trapped by the density cavities associated with IAWs.
Such phenomenon can occur frequently in plasmas. The interaction is then said
to be in a `strong turbulence' regime, in which transfer or redistribution of
energy to higher harmonic modes with small wavelengths may be possible. In
general, it is believed that some sort of chaotic process may be responsible
for the energy transfer from large to small spatial scales
\cite{Pettini,Tsaur}. This energy transfer may become faster when the chaotic
process is well developed in a subsystem of the full set of equations
\cite{Rizzato}.

One of the most important models in this context is the so-called Zakharov
equations (ZEs) \cite{Zakharov}, which have been derived by assuming a
quasineutrality as well as neglecting the ion-wave nonlinearity (IWN) due to
density fluctuation. However, the effects of dispersion due to charge
separation (deviation from the quasineutrality) may be important in the
radiation of IAWs, i.e., when the LWs \ resonantly interact with IAWs at the
same speed \cite{Karpman1,Karpman2}. \ On the other hand, when wave amplitude
becomes relatively large, the IWN can be important in the enhancement of
ion-acoustic wave emission as well as in the excitation of many unstable
harmonic modes at the initial stage of interaction, which later turns out to
spatiotemporal chaos (STC).\par
The basic purpose of this work is to investigate numerically the dynamics of
coupled LWs and IAWs in presence of these important effects, and also to
modify the results of Langmuir solitons not observed in the one-dimensional
(1D) ZEs. We show that transfer of energy to fewer modes, indeed, occurs and
it is faster when the chaotic process in a low-dimensional model of the full
system is well developed by the IWN. Such low-dimensional model is constructed
on the basis of three unstable harmonic modes \cite{Misra3}, and which depends
on a particular regime of the wave number of excitation. Numerical simulation
of this low-dimensional model reveals that transition from order to chaos is
possible, when the largest length scale of the IAWs is greater than that of
the soliton characteristic size. Solitons will then much be distorted by the
IAW emission until a mostly chaotic state emerges. For details of the idea of
constructing such model, readers are referred to the work in Ref.
\cite{Rizzato}.\par
Since the low-dimensional model is restricted to a few number of modes, and
also it is valid for a certain vale of the wave number of excitation, one
needs to investigate, e.g., numerically the the full space-time problem,
especially when this wave number is small enough to excite many unstable
modes. We show that these solitary patterns, thus formed, ultimate lose their
strengths after a long time through the\ random collision and fusion under the
strong IAW emission. The STC state of the system is then said to emerge.
However, the mechanism leading to such STC\ is still not clear. A few works in
this direction can be found in the literature \cite{Misra2,Rizzato,He}.\par
The paper is organized as follows. In Sec. II, we present the governing
equations, and analyze the evolution of Langmuir solitons in some particular
cases of interest. A low-dimensional three-wave model is formed and its linear
stability analysis is carried out in Sec. III. Numerical results for the
temporal evolution of the low-dimensional model and the spatiotemporal
evolution of the full system are presented in Sec. IV and Sec. V respectively.
Finally, Sec. VI contains some discussion and conclusion.
\section{Governing equations and the Langmuir soliton}
The nonlinear interaction of hf LWs and lf IAWs is governed by the following
coupled set\ of equations \cite{Karpman1,Karpman2}
\begin{align}
i\epsilon\frac{\partial E}{\partial t}+\frac{3}{2}\frac{\partial^{2}%
E}{\partial x^{2}}  &  =\frac{\nu}{2}E,\label{e1}\\
\frac{\partial^{2}\nu}{\partial t^{2}}-\frac{\partial^{2}\nu}{\partial x^{2}%
}-\frac{\partial^{4}\nu}{\partial x^{2}\partial t^{2}}  &  =\frac{\partial
^{2}|E|^{2}}{\partial x^{2}}\nonumber
\end{align}%
\begin{equation}
+\frac{\partial^{4}|E|^{2}}{\partial x^{2}\partial t^{2}}-\frac{1}{2}%
\frac{\partial^{2}\nu^{2}}{\partial x^{2}}-\frac{1}{2}\frac{\partial^{4}%
\nu^{2}}{\partial x^{2}\partial t^{2}}, \label{e2}%
\end{equation}
where $E(x,t)$ is the slowly varying Langmuir wave electric field normalized
by $\sqrt{16\pi n_{0}k_{B}T_{e}},$ with $n_{0}$ denoting the equilibrium
number density, $k_{B}$ is the Boltzmann constant, $\epsilon\equiv\sqrt
{m_{e}/m_{i}}$ is the square root of the electron-ion mass ratio and
$\nu(x,t)\equiv n_{e1}/n_{0}=n_{e}-1$ is the plasma relative density
fluctuation. The time $\ t$ is normalized by the ion plasma period
$\omega_{pi}^{-1}\equiv1/\sqrt{4\pi n_{0}e^{2}/m_{i}}$ and the space $x$ by
the electron Debye radius $\lambda_{De}\equiv\sqrt{k_{B}T_{e}/4\pi n_{0}e^{2}%
}.$ In Eq. (\ref{e2}) the terms with fourth-order mixed derivatives describe
the effects due to charge separation (i.e., deviation from the
quasineutrality). The third and fourth terms in the right-hand side of Eq.
(\ref{e2}) arise due to the expansion of $\log(1+\nu)$\ up to the second order
of $\nu$ in the equation for lf perturbation of electrons$.$ This enforces us
to assume that the relative density fluctuation associated with IAWs is not so
small$.$

We note that the analytic treatment of the Eqs. (\ref{e1}) and (\ref{e2})
without IWN has been studied in detail for the modification of Langmuir
solitons by Karpman \textit{et al} \cite{Karpman1,Karpman2}. However, when the
IWN is included in the system, the soliton properties may be changed. \ To
look for this effect, we simply disregard the mixed derivatives, and solve
numerically the system of Eqs. (\ref{e1}) and (\ref{e2}) by Runge-Kutta
scheme. In the scheme, we approximate the spatial derivatives by the centered
second-order difference formulas, and use $\nu=E=\partial^{2}E/\partial
x^{2}=\partial^{2}\nu/\partial x^{2}=0$ at the boundaries$.$ The results are
presented in Fig.1. When the ion-acoustic radiation is weak, we see that
around $x=0$ the ponderomotive force dominates over the IWN, giving rise an
increased amplitude of the envelope (highly correlated with the density
depletion). However, in the other regimes of $x$, IWN dominates and makes the
solitons wider. Later, we will see that as time progresses, the IAW emission
increases by the IWN, causing solitons to lose their strengths.
\begin{figure}[ptb]
\begin{center}
\includegraphics[height=3.5in,width=3.5in]{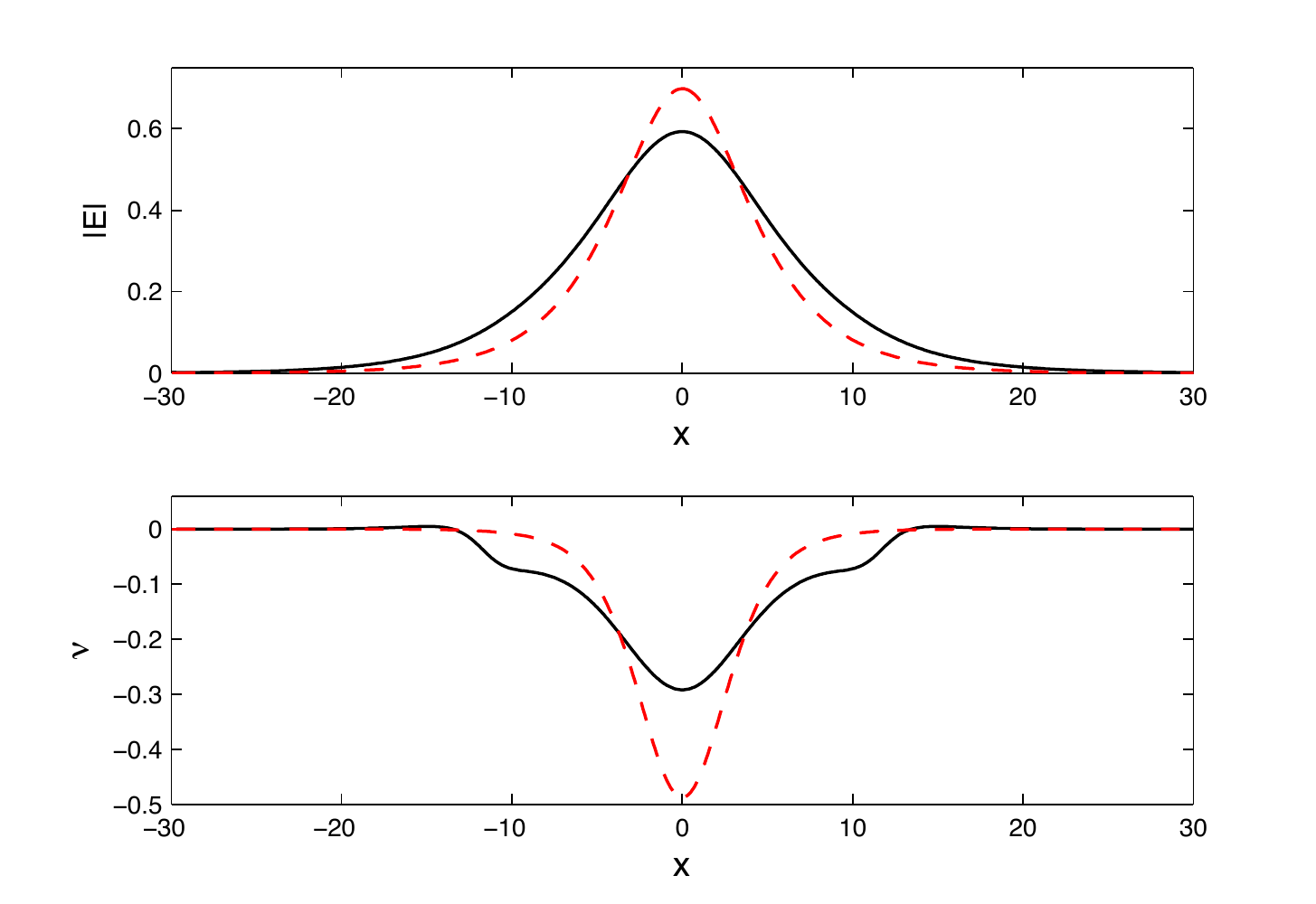}
\end{center}
\caption{(Color online) The profiles of the Langmuir wave electric field
(upper panel) and the associated density depletion (lower panel) are shown
[solution of Eqs. (1) and (2)]. The solid and dashed (red) lines respectively
correspond to the case with IWN and without IWN after $t=10(\omega_{pi}^{-1}%
)$.}%
\end{figure}Some particular cases can be of interest in the context of
Langmuir solitons. For example, under the small amplitude condition, i.e.,
disregarding the third \ and fourth terms in the right-hand side of Eq.
(\ref{e2}), and assuming the quasineutrality (disregarding the mixed fourth
order derivatives), the dynamics of small amplitude IAWs are then governed by%

\begin{equation}
\frac{\partial^{2}\nu}{\partial t^{2}}-\frac{\partial^{2}\nu}{\partial x^{2}%
}=\frac{\partial^{2}|E|^{2}}{\partial x^{2}}. \label{e3}%
\end{equation}
This equation when coupled with Eq. (\ref{e1}) describes under suitable
renormalizations the classical ZEs \cite{Zakharov}, which has been extensively
studied in the context of Langmuir wave turbulence (see e.g.,
\cite{Hadzievski,Doolen,Prado,Misra2,Misra3,Chain}).

Again, assuming a slow space-time response to the perturbations as
$x\rightarrow\epsilon^{1/2}(x-t),t\rightarrow\epsilon^{3/2}t,$ i.e., ion
perturbations move with the ion-acoustic speed, $c_{s}\equiv\sqrt{k_{B}%
T_{e}/m_{i}}$, and asuming\ the dependent variables to vary$\sim\epsilon
,$\ the IAWs can be described by the following modified Korteweg-de Vries
(mKdV) equation \cite{Makhankov,Nishikawa,Ikezi}%

\begin{equation}
2\frac{\partial\nu}{\partial t}-\frac{1}{2}\frac{\partial\nu^{2}}{\partial
x}+\frac{\partial^{3}\nu}{\partial x^{3}}+\frac{\partial|E|^{2}}{\partial
x}=0. \label{e4}%
\end{equation}
The equation (\ref{e1}) together with Eq. (\ref{e4}) describe the dynamics of
slowly varying Langmuir wave electric field coupled to the slow response of
the density perturbations. These coupled equations predict a class of
solutions with a bell-shaped Langmuir wave electric field profile arrested
into the density cavity \cite{Nishikawa,Ikezi}.

We now turn out to the global behaviors of the system. To this end, we first
renormalize the original variables (similar to the ZEs) as $x\rightarrow
\epsilon x,t\rightarrow\epsilon t,\nu\rightarrow\nu/\epsilon^{2},E\rightarrow
E/\epsilon,$ and obtain the condition for MI as given below. Thus, Eqs.
(\ref{e1}) and (\ref{e2}) reduce to%

\begin{align}
2i\frac{\partial E}{\partial t}+3\frac{\partial^{2}E}{\partial x^{2}}  &  =\nu
E\label{e5}\\
\frac{\partial^{2}\nu}{\partial t^{2}}-\frac{\partial^{2}\nu}{\partial x^{2}%
}-\epsilon^{2}\frac{\partial^{4}\nu}{\partial x^{2}\partial t^{2}}  &
=\frac{\partial^{2}|E|^{2}}{\partial x^{2}}\nonumber
\end{align}%
\begin{equation}
+\epsilon^{2}\frac{\partial^{4}|E|^{2}}{\partial x^{2}\partial t^{2}}%
-\frac{\epsilon^{2}}{2}\frac{\partial^{2}\nu^{2}}{\partial x^{2}}%
-\frac{\epsilon^{4}}{2}\frac{\partial^{4}\nu^{2}}{\partial x^{2}\partial
t^{2}} \label{eq6}%
\end{equation}
Clearly, the term proportional to $\epsilon^{4}(\sim10^{-7})$ becomes smaller
and can be neglected. However, the terms proportional to $\epsilon^{2}$ can no
longer be neglected when the soliton attenuation becomes relatively larger.

Next, the linear stability analysis of the perturbations $\propto$
$\exp(ikx-i\omega t)$ [where $k$ $(\omega)$ is the wave number (frequency) of
modulation] for Eqs. (\ref{e5}) and (\ref{eq6}) gives the following growth
rate of MI.%

\begin{equation}
\gamma=\frac{k}{\sqrt{2\alpha}}\left[  \sqrt{\Lambda^{2}+6\alpha\left(
|E_{0}|^{2}-\frac{3k^{2}}{2}\right)  }-\Lambda\right]  ^{1/2}, \label{e7}%
\end{equation}
where $\alpha=1+k^{2}\epsilon^{2},$ $\Lambda=1+3k^{2}\left(  3\alpha
/2-\epsilon^{2}|E_{0}|^{2}\right)  /2,$ and the MI sets in for $k$ satisfying
$0<k<k_{c}\equiv\sqrt{2/3}|E_{0}|.$

\section{Low-dimensional (three-wave) model}

We note that the system of Eqs. (\ref{e5}) and (\ref{eq6}) is, in general,
multi-dimensional. However, there may be the situation in which a few number
of modes is more active than the remaining modes. Such cases are quite typical
not only for Langmuir wave turbulence, but are rather general for parametric
instabilities of hf waves interacting with lf ones close to the instability
threshold. \ In this way, Galerkin expansions and truncations to a few normal
modes are commonly used to describe the basic features of the full dynamics by
a low-dimensional model. However, we will see that the specific details of
such model are highly dependent on the range for the basic wave number of
modulation $k$. So, considering the dynamics of few coupled waves, we expand
the envelope $E(x,t)$ and the density $\nu(x,t)$ as \cite{Rizzato}%

\begin{align}
E(x,t)  &  =\sum_{m=-M/2}^{+M/2}E_{m}(t)e^{imkx}=\sum_{m=-M/2}^{+M/2}\rho
_{m}(t)e^{i\theta_{m}(t)}e^{imkx},\label{E}\\
\nu(x,t)  &  =\sum_{m=-M/2}^{+M/2}n_{m}(t)e^{imkx}, \label{n}%
\end{align}
where $M=[k^{-1}]$ represents the number of modes, $\rho_{m}=\rho_{-m}%
,\theta_{m}=\theta_{-m},n_{m}=n_{-m}$. Since in our case, $M=2,$ $\rho
_{1}=\rho_{\pm1},n_{1}=n_{\pm1}$ etc., we can express the fields as (see for
details \cite{Misra1,Misra3,Batra})
\begin{align}
E  &  =E_{0}+E_{1}\cos(kx)\equiv\sqrt{N}\sin\left(  \frac{a}{2}\right)
\exp(i\theta_{0})\nonumber\\
&  +\sqrt{2N}\cos\left(  \frac{a}{2}\right)  \exp(i\theta_{1})\cos
(kx)\,,\label{e8}\\
\nu &  =N+n_{1}\cos(kx)\,, \label{e9}%
\end{align}
where $N$ $(\equiv n_{0})=|E_{-1}|^{2}+|E_{0}|^{2}+|E_{1}|^{2}$ is the
conserved plasmon number. Also, $a,$ $\theta_{0},$ $\theta_{1}$ and $n_{1}$
are each a function of time $t$. \ Inserting Eqs. (\ref{e8}) and (\ref{e9})
into the Eqs. (\ref{e5}) and (\ref{eq6}), and following, e.g., Refs.
\cite{Misra1,Misra3,Batra} we obtain the following temporal system
\begin{equation}
\dot{a}=-\frac{n_{1}}{\sqrt{2}}\sin\varphi, \label{e10}%
\end{equation}%
\begin{equation}
\dot{\varphi}=\frac{1}{2}\left(  3k^{2}-\sqrt{2}n_{1}\cos\varphi\cot a\right)
\,, \label{e11}%
\end{equation}%
\begin{equation}
\ddot{n}_{1}=U_{1}n_{1}^{2}+U_{2}n_{1}+U_{3}\,, \label{e12}%
\end{equation}
where the `dot' represents differentiation with respect to $t,$ $\varphi
=\theta_{0}-\theta_{1},$ and $U_{1},$ $U_{2}$ and $U_{3}$ are given by
\begin{align}
U_{1}(a,\varphi)  &  =C_{1}\frac{\cos\varphi}{\sin a}\left(  \sin^{2}%
\varphi+\cos^{2}a\right)  ,\nonumber\\
U_{2}(a,\varphi)  &  =C_{2}+\left(  C_{3}-U_{1}\sin a\right)  \cos
a,\nonumber\\
U_{3}(a,\varphi)  &  =C_{4}\sin a\cos\varphi. \label{e13}%
\end{align}
Here the constants $C_{1},...,C_{4}$ are
\begin{align}
C_{1}  &  =\frac{Nk^{2}\epsilon^{2}}{\sqrt{2}(1+k^{2}\epsilon^{2})}%
,C_{2}=-\frac{k^{2}\left(  1-N\right)  }{1+k^{2}\epsilon^{2}},\nonumber\\
C_{3}  &  =-\frac{3k^{2}}{\sqrt{2}}C_{1},C_{4}=\frac{\sqrt{2}Nk^{2}}%
{1+k^{2}\epsilon^{2}}\left(  \frac{9k^{4}\epsilon^{2}}{4}-1\right)
\label{e14}%
\end{align}
The system of Eqs. (\ref{e10})-(\ref{e12}) \ can be recast as a set of four
ordinary differential equations (ODEs):%

\begin{align}
\dot{x}_{1}  &  =-\frac{1}{\sqrt{2}\epsilon}\,x_{3}\sin x_{2}\,,\label{eq15}\\
\dot{x}_{2}  &  =\frac{1}{2\epsilon}\left(  3k^{2}-\sqrt{2}x_{3}\,\cos
x_{2}\,\cot x_{1}\right)  \,,\label{e16}\\
\dot{x}_{3}  &  =x_{4}\,,\label{e17}\\
\dot{x}_{4}  &  =U_{1}x_{3}^{2}+U_{2}x_{3}+U_{3}\,, \label{e18}%
\end{align}
where $U_{1},U_{2},U_{3}$ are now functions of $x_{1},$ $x_{2},$ and we
redefine the variables as $a=x_{1},\varphi=x_{2},n_{1}=x_{3},\dot{n}_{1}%
=x_{4}.$ The basic properties of the system can now be in order. It can be
shown that the adiabatic limit of Eqs. (\ref{eq15})-(\ref{e18}) gives an
integrable system, and hence nonchaotic \cite{Misra3,Batra,Rizzato}. Next, the
local stability of the system about the fixed points $(x_{10},x_{20}%
,x_{30},0)$ given by%

\begin{equation}
\sin x_{20}=0,\cot x_{10}=\pm\frac{3k^{2}}{\sqrt{2}x_{30}}, \label{e19}%
\end{equation}

\[
\pm3C_{3}k^{2}\pm\sqrt{2}C_{4}+C_{2}\sqrt{9k^{4}+2x_{30}^{2}}%
\]%
\begin{equation}
+\frac{9C_{1}k^{4}}{\sqrt{2}}\left(  \pm1-\frac{3\sqrt{2}k^{2}}{9k^{4}%
+2x_{30}^{2}}\right)  =0, \label{e20}%
\end{equation}
where $x_{30}\neq0,\cos x_{20}=\pm1,$ can be examined by linearizing the Eqs.
(\ref{eq15})-(\ref{e18}), and solving the corresponding eigenvalue problem.
The linearized set of equations in a neighborhood of $Q\equiv(x_{10}%
,x_{20},x_{30},0)$ is%

\begin{equation}
\dot{X}=Df(Q)X, \label{e21}%
\end{equation}
where $X$ $=[x_{1},x_{2},x_{3},x_{4}]$ is a column vector and $Df(Q)$ is the
Jacobian matrix at $Q$, i.e.,%

\begin{equation}
Df(Q)=\left(
\begin{array}
[c]{cccc}%
0 & -p_{0} & 0 & 0\\
q_{0} & 0 & -r_{0} & 0\\
0 & 0 & 0 & 1\\
A_{1} & A_{2} & A_{3} & 0
\end{array}
\right)  , \label{e22}%
\end{equation}
The eigenvalue equation $|Df(Q)-\lambda I|=0$ then reads%

\[
\lambda^{4}+\left(  A_{3}+p_{0}q_{0}\right)  \lambda^{2}+\left(  A_{2}%
r_{0}\right)  \lambda
\]

\begin{equation}
+p_{0}\left(  A_{3}q_{0}-A_{1}r_{0}\right)  =0, \label{e23}%
\end{equation}
where
\begin{equation}
p_{0}=\pm\frac{x_{30}}{\sqrt{2}\epsilon},q_{0}=\frac{p_{0}}{\sin^{2}x_{10}%
},r_{0}=\pm\frac{\cot x_{10}}{\sqrt{2}\epsilon}, \label{e24}%
\end{equation}

\begin{align}
A_{1}  &  =-C_{3}\sin x_{10}\pm C_{4}\cos x_{10}\nonumber\\
&  \pm C_{1}\cos^{2}x_{10}\left(  1-\cos x_{10}\cot x_{10}\right) \nonumber\\
&  \pm\left(  C_{1}\cos x_{10}-\frac{x_{30}}{\sin x_{10}}\right) \nonumber\\
&  \times\left(  \sin(2x_{10})+\cot x_{10}\cos^{2}x_{10}\right)  , \label{e25}%
\end{align}

\begin{equation}
A_{2}=\pm2x_{30}\cos x_{10},A_{3}=A_{2}\cot x_{10}.\label{e26}%
\end{equation}
We numerically examine a set of fixed points satisfying Eqs. (\ref{e19}) and
(\ref{e20}), and the linear stability of the system by the Routh-Hurwitz (RH)
criteria. To this end, we use Carden's method to look for a real solution of
Eq. (\ref{e20}) when expressed it as a cubic in $\alpha\equiv\sqrt{2x_{30}%
^{2}+9k^{4}\text{ }}$. The range of $k$ and $N$ for such real $\alpha,$ and
hence for $x_{30}$ are $0<k\leq0.5<\sqrt{2/3}|E_{0}|$ and $0.6\leq N\leq8.$
Thus, for a given set of values of $k$ and $N$, one can obtain a set of fixed
points from Eqs. (\ref{e19}) and (\ref{e20}). Having obtained a set of fixed
points, one can then numerically examine the coefficients of the depressed
quartic [Eq. (\ref{e23})] for $0<k\leq0.5$ and $0.6\leq N\leq8$. We find that
in these regimes, coefficients of $\lambda^{2}$ and $\lambda$ are always
positive, and the constant term changes sign as in Fig. 2. Here the shaded
(white) region indicates the positive (negative) values of the constant term.
The constant term is also positive for $0<k<0.49$ and $0.6\leq N\leq8.$ Thus,
by the RH criteria Eq. (\ref{e23}) has at least one root with positive real
part when the constant term of the quartic is negative. The system of Eqs.
(\ref{eq15})-(\ref{e18}) is then said to be linearly unstable. \ As an
illustration, the eigenvalues for $N=8$, $0.498\leq k\leq0.5,$ are a pair of
complex conjugates, one real positive $(>1)$ and one negative real roots.
\begin{figure}[ptb]
\begin{center}
\includegraphics[height=2.5in,width=3in]{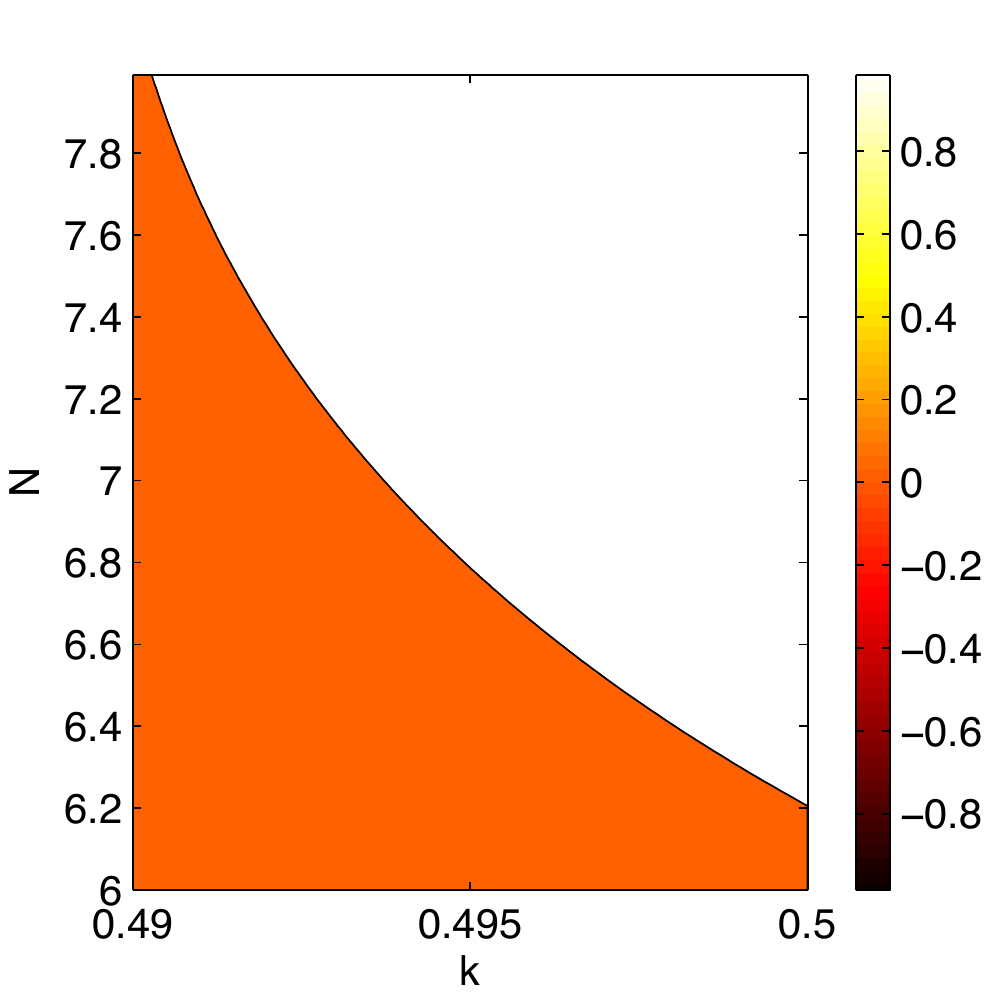}
\end{center}
\caption{(Color online) The plot of the constant coefficient in the eigenvalue
equation (25) with the variations of $N$ and $k$, showing its positive (shaded
colored region) and negative (white region) values. The coefficient is also
positive (not shown) for \ $0<k<0.49$ and $0.6\leq N\leq8.$}%
\end{figure}

\section{Numerical simulation for a three-wave model: the temporal dynamics}

From Eq. (\ref{e7}), the MI sets in only if $k<k_{c}\equiv\sqrt{2/3}|E_{0}|,$
and\ $k=k_{c}$ defines the curve along which pitchfork bifurcation takes place
\cite{Oliveria}. Also, the dynamics is really subsonic in $k_{c}/2<k<k_{c}$
where the MI growth rate is small \cite{Rizzato}. As, $k$ decreases from
$k_{c}/2,$ many more unstable modes (since $M=[k^{-1}])$ with higher harmonics
will be excited. In this case, even though the dynamics is not subsonic, a
description with three modes should be relatively accurate for $k<k_{c}/2.$
Our basic assumption at this point is that the Langmuir modes for $|m|>1$ are
stable, and the ion-acoustic modes with $|m|=2$ remain as they are, already
excited by the presence of $|m|=1$ Langmuir wave envelopes. This results a
system of ODEs [Eqs. (\ref{eq15})-(\ref{e18})] describing the temporal
dynamics of the field variables. Notice, however, that although the Fourier
modes are truncated, no assumption on the slowness of the dynamics is made. We
would like to see how the temporal system behaves as $k$ approaches $k_{c}$
from $k_{c}/2,$ and the condition for the subsonic region is progressively
relaxed by reducing $k$ from $k_{c}/2$ to some extent$.$ Here $k$ is not so
small that the three-wave model becomes inappropriate, because smaller the
$k$, the larger is the number of unstable modes. In the latter case, full
simulation of Eqs. (\ref{e5}) and (\ref{eq6}) will be necessary to give more
appropriate information about the interaction.

Thus, for the temporal evolution, we numerically integrate the system of Eqs.
(\ref{eq15})-(\ref{e18}) by\ the Runge-Kutta scheme with step size $h=0.001$
and initial values $x_{1}=0.1,x_{2}=0.1,x_{3}=0.2,x_{4}=0.3.$ By suitably
varying the system parameters $k$ and $N$ together with the initial choices,
we have obtained different sets of parameters which can exhibit periodic as
well as chaotic attractors. The\ latter are established by the Lyapunov
analysis with at least one positive exponent indicating chaos.\ On the other
hand, the regular or limit cycles may be described on the basis of their
structures, amplitude oscillations with time or Lyapunov exponents with
negative or zero values.

For $N=6$, $E_{0}=2$ and $k=1.5$ in $k_{c}/2<k<k_{c},$ Fig. 3 shows a periodic
limit cycle. This suggests that within the plane wave region, strong chaotic
activity is completely absent, i.e., most of the Kolmogorov-Arnol'd-Moser
(KAM) surfaces, especially the central fixed point remains unaffected by the
nonlinear interaction. In fact, the absence of chaos is not only for these
parameter values, but also for some other values smaller or larger. It also
explains that within this\ plane wave region no massive redistribution of
energy takes place. This information well agrees with the full simulation of
the space-time problem \cite{Rizzato}. \ As $k$ is lowered from $k_{c}/2$,
e.g., $k=0.5$, Fig. 4\textbf{ }shows that the periodicity or integrability of
the temporal system tends to become poorer exhibiting chaotic orbits. \ This
can be verified by the set of corresponding Lyapunov exponents as shown in
Fig. 5, where one positive exponent indicates chaos. \ We\ also find that as
$k$ becomes smaller, the largest Lyapunov tends to increase from a positive
small value to a large one $(>1)$. So, it may be possible that as the basic
wave number $k$ \ is decreased further, strong chaotic features develop into
the system. \begin{figure}[ptb]
\begin{center}
\includegraphics[height=2.5in,width=4.0in]{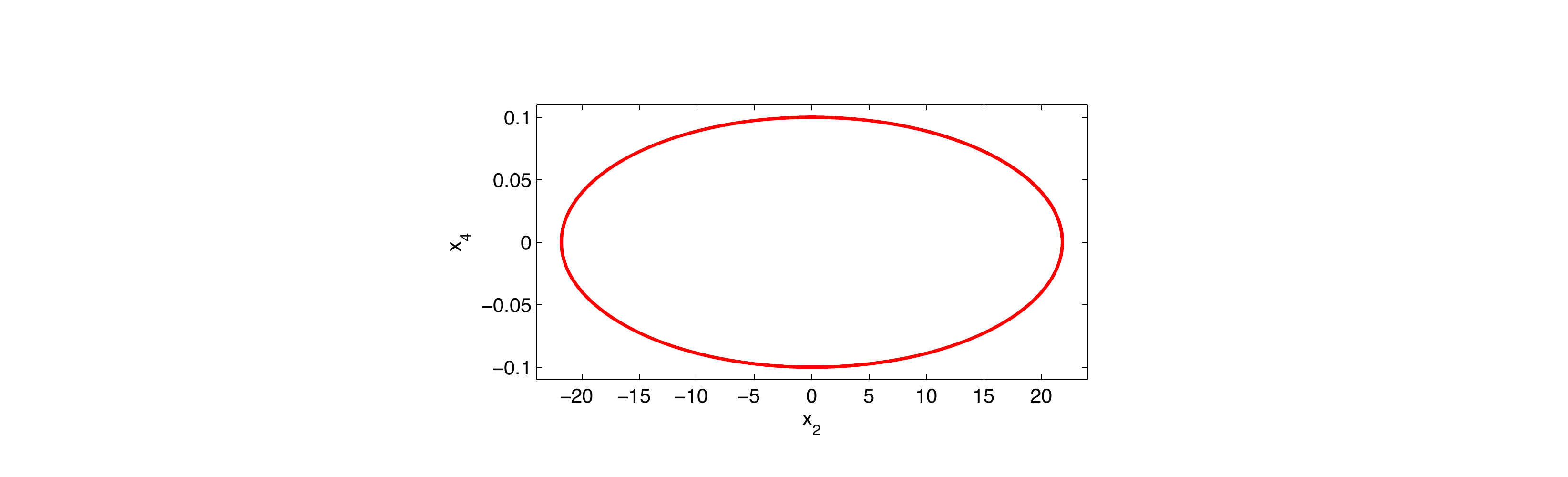}
\end{center}
\caption{(Color online) A periodic limit cycle in the low-dimensional model
for $k=1.5,N=6$ and $E_{0}=2$, showing that the three-wave dynamics is
stable.}%
\end{figure}\begin{figure}[ptb]
\par
\begin{center}
\includegraphics[height=2.5in,width=4.0in]{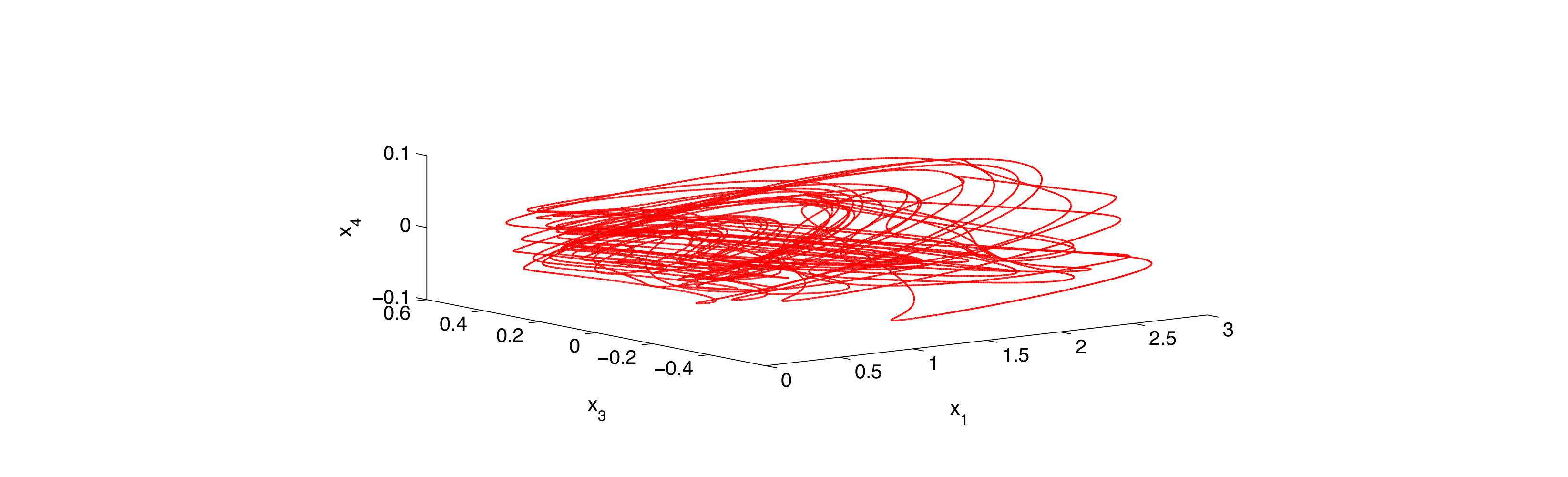}
\end{center}
\caption{(Color online) A chaotic phase portrait in the low-dimensional model
for $k=0.5,N=6$ and $E_{0}=2$, showing that the three-wave dynamics is
unstable.}%
\end{figure}\begin{figure}[ptb]
\begin{center}
\includegraphics[height=2.5in,width=4.0in]{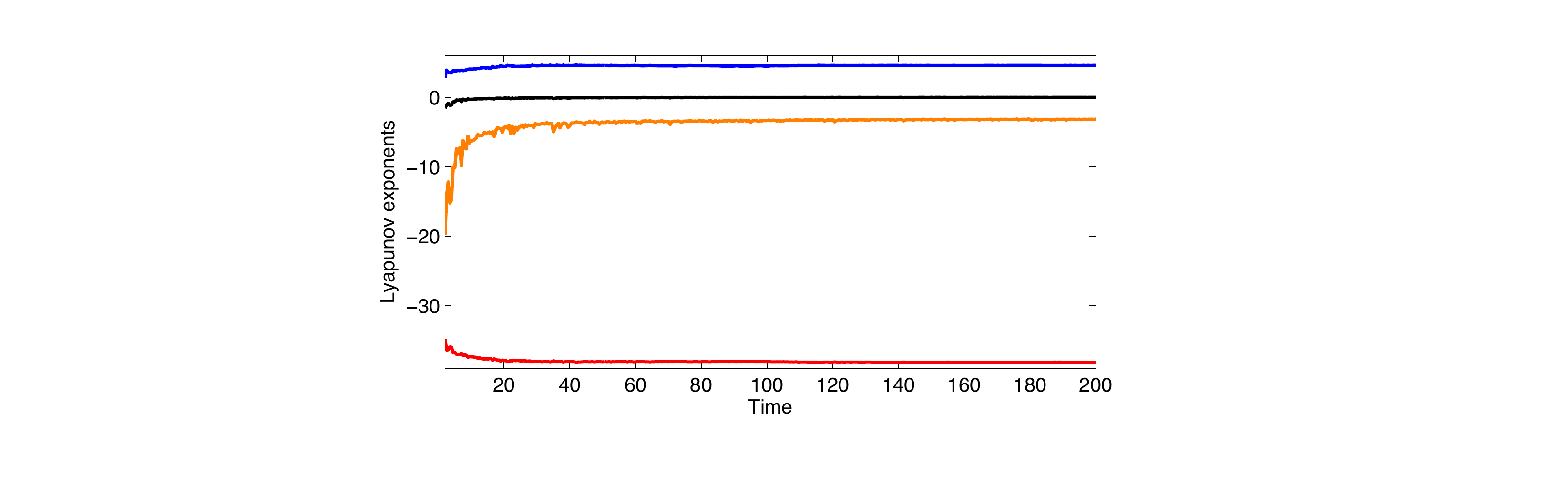}
\end{center}
\caption{(Color online) The four Lyapunov exponents with respect to time
$(t\epsilon)$ corresponding to the same parameter values as in Fig. 4. One of
the exponents is clearly positive, indicating that the three-wave dynamics is
chaotic.}%
\end{figure}\begin{figure}[ptb]
\begin{center}
\includegraphics[height=2.5in,width=4in]{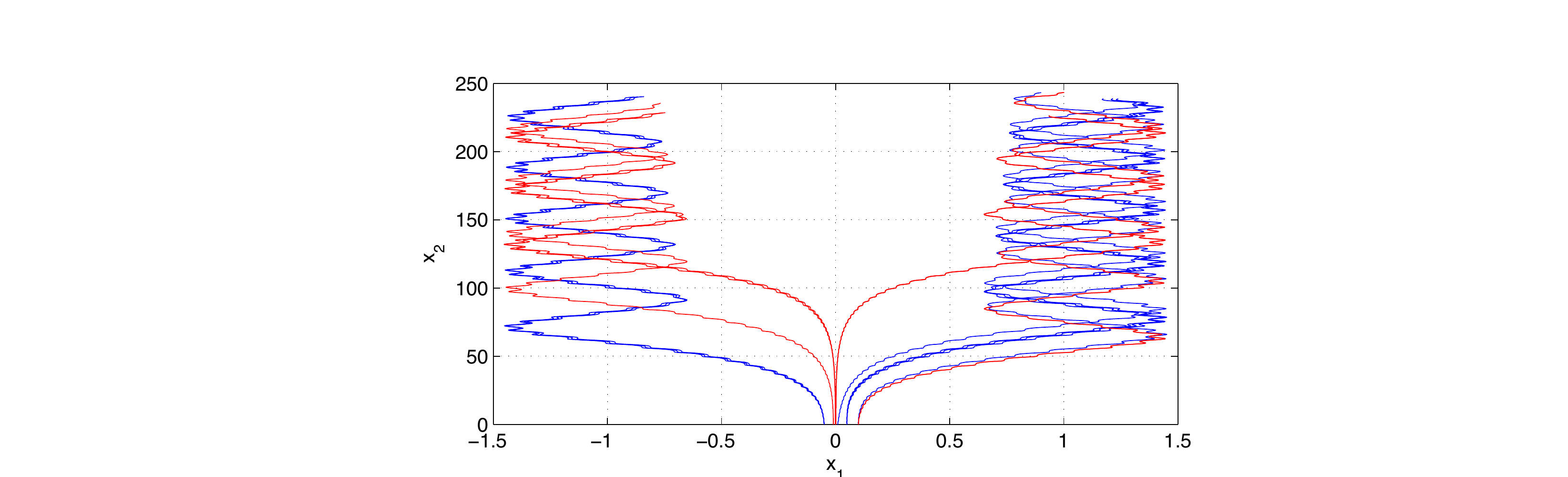}
\end{center}
\caption{(Color online) Supercritical Hopf-bifurcation in the low-dimensional
model for $k=0.45,N=7$ and $E_{0}=2$, showing the bifurcation of an attracting
periodic orbit.}%
\end{figure}Thus, when $k<k_{c}/2,$ the system's periodicity breaks down. This
is expected, since for smaller $k$, the solitons get more affected by the
ion-acoustic radiation until they are completely destroyed. In the strong
chaotic process one can expect that solitons will deliver their energy to the
remaining modes so as to gradually \ disappear after some time. However, one
can no longer rely on the low-dimensional model for some smaller values of
$k$, as it gives rise an excitation of large number of unstable harmonic
modes. This is the situation where full simulation of Eqs. (\ref{e5}) and
(\ref{eq6}) will give better understanding for the interaction.

In addition to the above results, one interesting can also be to study the
existence of Hopf-bifurcation in the Eqs. (\ref{eq15})-(\ref{e18}) by choosing
$k$ as the bifurcation parameter. Generally, Hopf-bifurcation is defined as
the change in qualitative behavior of the system when a pair of complex
conjugate eigenvalues passes through the imaginary axis. A pair of such roots
can be obtained by numerically examining the eigenvalue equation (\ref{e23}).
\ Thus, one can perform a series of analysis with different values of
$k$.\ Figure 6 shows the supercritical diagram corresponding to $N=7$ and
$k=0.45$ in which an attracting periodic orbit bifurcates. The detailed
analysis of such bifurcation is, however, beyond the scope of the present work.

\section{Simulation results for the spatiotemporal evolution}

In the previous section we have observed that as the value of $k$ is lowered
from $k_{c}/2,$ the system exhibits strong chaotic features. So, in order to
investigate the global behaviors of the solitons, namely the soliton
interaction with strong IAW emission, formation of incoherent patterns due to
collision and fusion, it is reasonable to consider the system of Eqs.
(\ref{e5}) and (\ref{eq6}) without the mixed derivatives (valid for
long-wavelengths). In our numerical scheme, we assume the spatial periodicity
with the simulation box length $L_{x}$ such that $L_{x}=2\pi/k,$ the resonant
wave length$.$ The integer $M=[k^{-1}],$ representing the number of modes, is
considered as a number of grid points, e.g., $256,512,1024$ and $2048$ as the
case may be. We choose the initial conditions as \cite{Misra2,He}%
\begin{align}
E(x,0)  &  =E_{0}\left[  1+b\cos(kx)/L\right]  ,\nonumber\\
\text{ }n(x,0)  &  =-2bE_{0}\cos(kx)/L, \label{e27}%
\end{align}
where $E_{0}$ is the amplitude of the pump Langmuir wave, $b$ is a\ suitable
constant, and $L$ is of the order of $10^{-3}$ to emphasize that the
perturbation is relatively small. The spatiotemporal system was advanced in
time using the Runge-Kutta scheme with a\ time step $dt=0.001$. We consider
the grid size $M=2048$ for lower values of $k$, and $M=1024$ for relatively
large $k,$ so that $x=0$ corresponds to the grid position $1024$ and $512$
respectively. In the simulation, we consider $E_{0}=2,$ $b=1\ $and
$\epsilon=\sqrt{1/1840}$, unless otherwise mentioned.

Figures 7 (in presence of IWN) and 8 (in absence of IWN) show the profiles of
the electric filed (upper panel) and density fluctuation (lower panel)
associated with the IAWs at the end of simulation for $k=0.22,$ i.e., for the
of $M\equiv\lbrack k^{-1}]=4$ harmonic modes. At the central part, we observe
an excited electric field, $|E|\sim9.5\epsilon$ highly correlated with the
density depletion, $\nu=-81.2\epsilon^{2}$ in presence of IWN. Whereas in
absence of IWN, these are $|E|\sim9.375\epsilon$ and $\nu=-84.4\epsilon^{2}.$
Clearly, the effect of IWN is to enhance the magnitude of the wave electric
field or to decrease that of the density fluctuation. \ From Fig. 9 (contour
plot corresponding to Fig. 7), we observe that the solitary master patterns
are first formed by the master mode, and then begin to appear from the
unstable harmonic modes. During the pattern formation, ions are\ usually
driven out (by the resultant of ponderomotive force and the IWN) from the
regions of the master mode as well as from some harmonic patterns, and then
they form density distributions with $\nu>0$ (see the lower panels of Figs. 7
and 8) along the spatial axis. These ion density humps move stochastically on
either side of the patterns, and do not arrest the Langmuir wave fields. We
observe from Figs. 9 and 10\textbf{ }(contour plot corresponding to Fig. 8)
that the pattern selection leads to three $(<M=4)$ harmonic modes initially
peaked at $x=L_{x}/4\epsilon,L_{x}/2\epsilon$ and $3L_{x}/4\epsilon.$ These
three modes are basically generated by the master mode and the unstable
harmonic modes. The system is then in the coexistence of temporal chaos (TC)
and spatio-partial coherence (SPC) state similar to the case\ of plane wave
region $(k_{c}/2<k<k_{c})$ explained in the previous section.

However, as the IWN is relaxed, only two solitary patterns initially peaked at
$x=L_{x}/4\epsilon$ and $3L_{x}/4\epsilon$ are seen to collide with the master
mode at $t\approx120/\epsilon$ and $t\approx170/\epsilon$ respectively (see
Fig. 10)$.$ Also, the collision is so weak that it does not influence the
soliton dynamics. The reason is that in absence of IWN and for relatively
small $E_{0}$, the ponderomotive force was the only force to cause the weak
ion-acoustic wave emission. As a result, though the collisions take place but
are not sufficiently strong to fuse into a new one. On the other hand, when
$E_{0}$ is\ relatively large, e.g., $E_{0}=4,$ i.e.,\ when the amplitude of
both the initial profiles increase, one may observe (see Fig. 11) more
collision and fusion due to strong IAW emission by the IWN. In this case,
$|E|$ increases, but $|\nu|$ decreases as compared to the case of no IWN (Fig.
12). \ Figure 11 also shows that two pairs of solitary patterns initially
peaked at the either side of $x\approx256/\epsilon,$ collide and fuse into two
new patterns, one of which again collides with the other two master patterns
peaked at $x\approx256/\epsilon$ and $x\approx465/\epsilon$ at $t\approx
115/\epsilon$ and $165/\epsilon.$ After these collisions, there remain only
three distorted patterns. The system is in TC, but still in the coexistence of
TC and STC, since a few collisions will not suffice to cause STC \cite{He}.
\begin{figure}[ptb]
\begin{center}
\includegraphics[height=3.5in,width=3.5in]{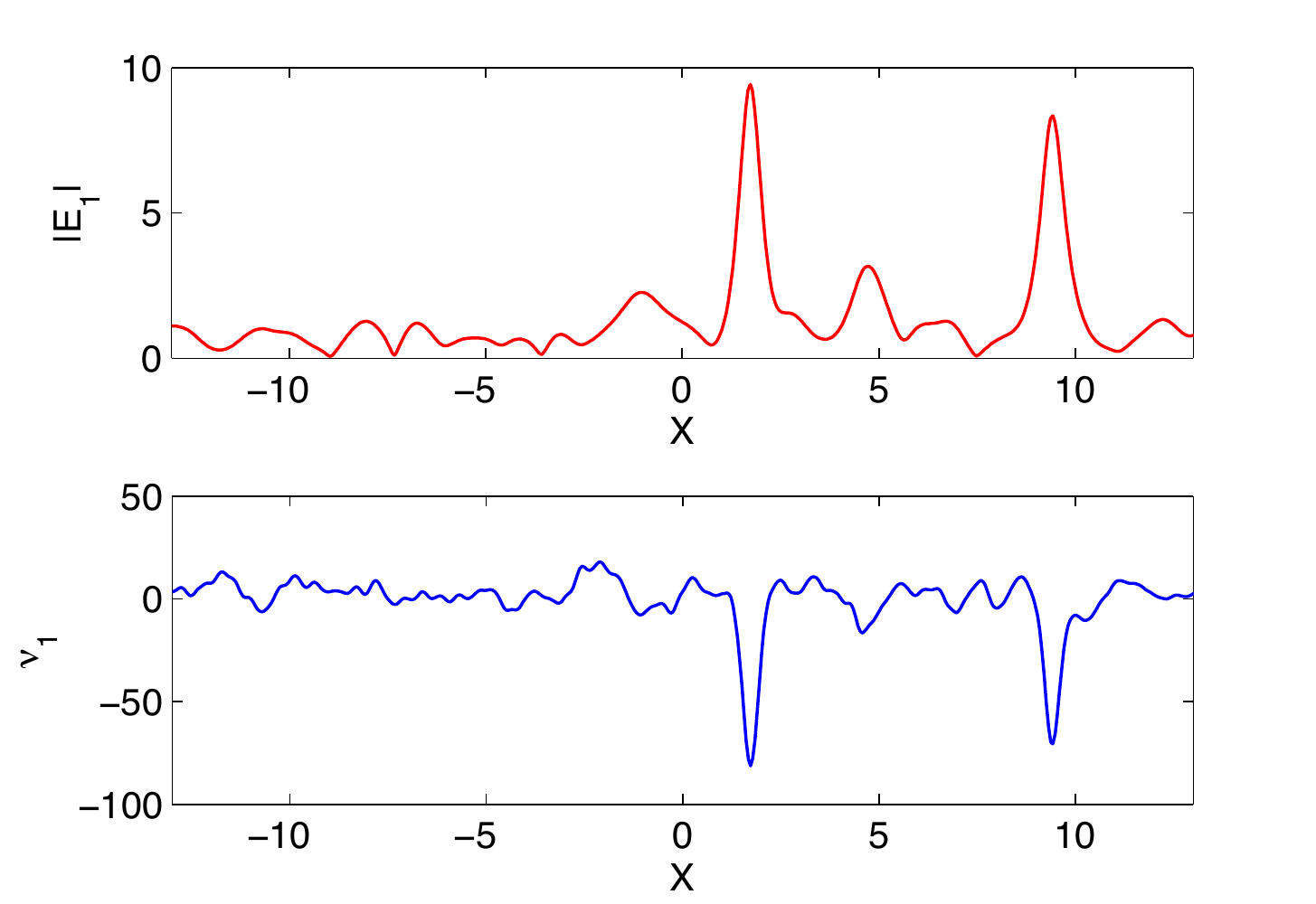}
\end{center}
\caption{(Color online) The profiles of the wave electric field ($|E_{1}%
|\equiv|E|/\epsilon,$ upper panel) and the associated density fluctuation
($\nu_{1}\equiv\nu/\epsilon^{2},$ lower panel) with respect to the space
$(X\equiv x\epsilon)$ after time $t=200/\epsilon$ in the numerical simulation
of Eqs. (5) and (6) for $k=0.22$ and $E_{0}=2$ with IWN.}%
\end{figure}\begin{figure}[ptb]
\begin{center}
\includegraphics[height=3.5in,width=3.5in]{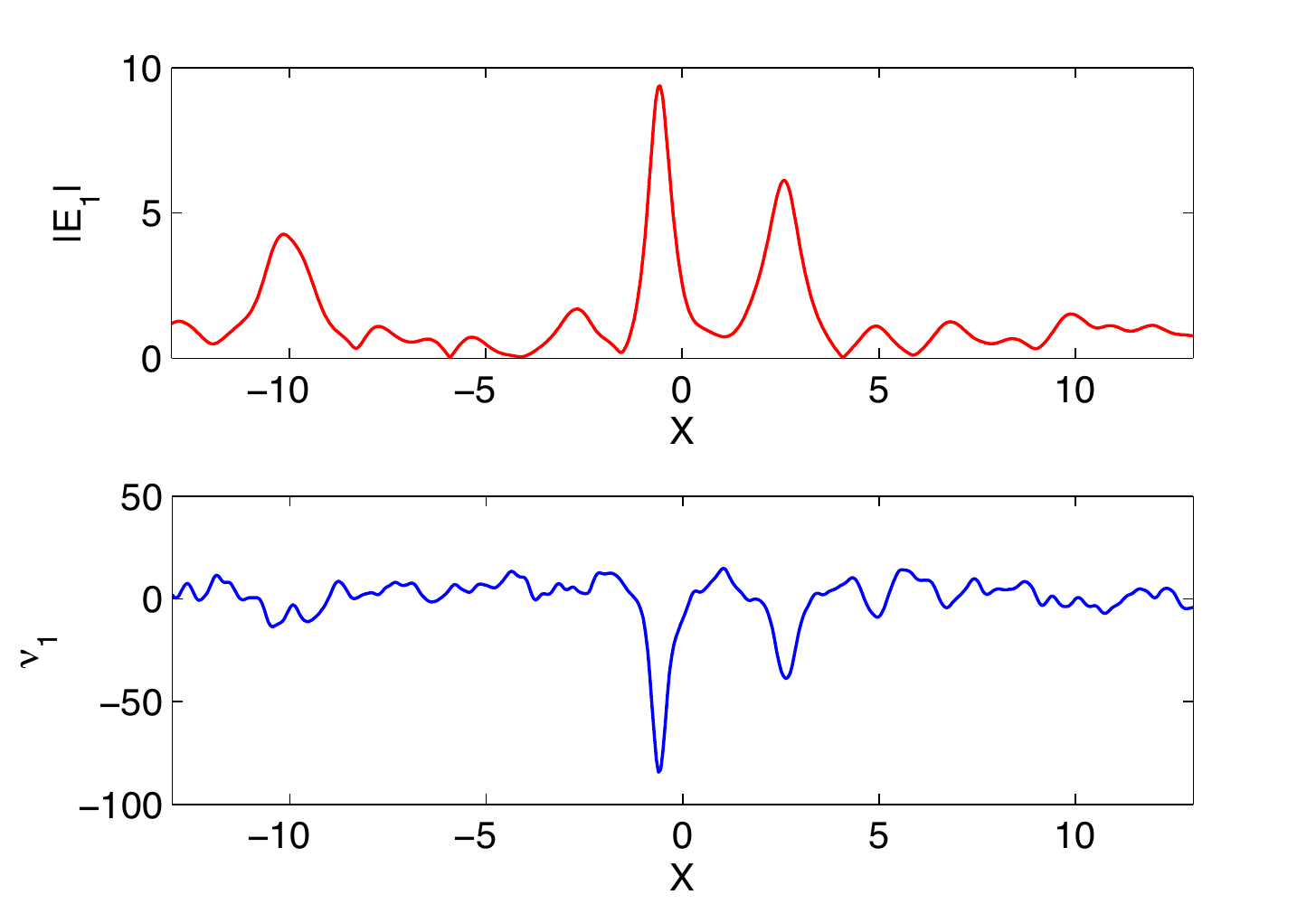}
\end{center}
\caption{(Color online) The same as in Fig. 7, but without IWN.}%
\end{figure}\begin{figure}[ptb]
\begin{center}
\includegraphics[height=2.5in,width=2.5in]{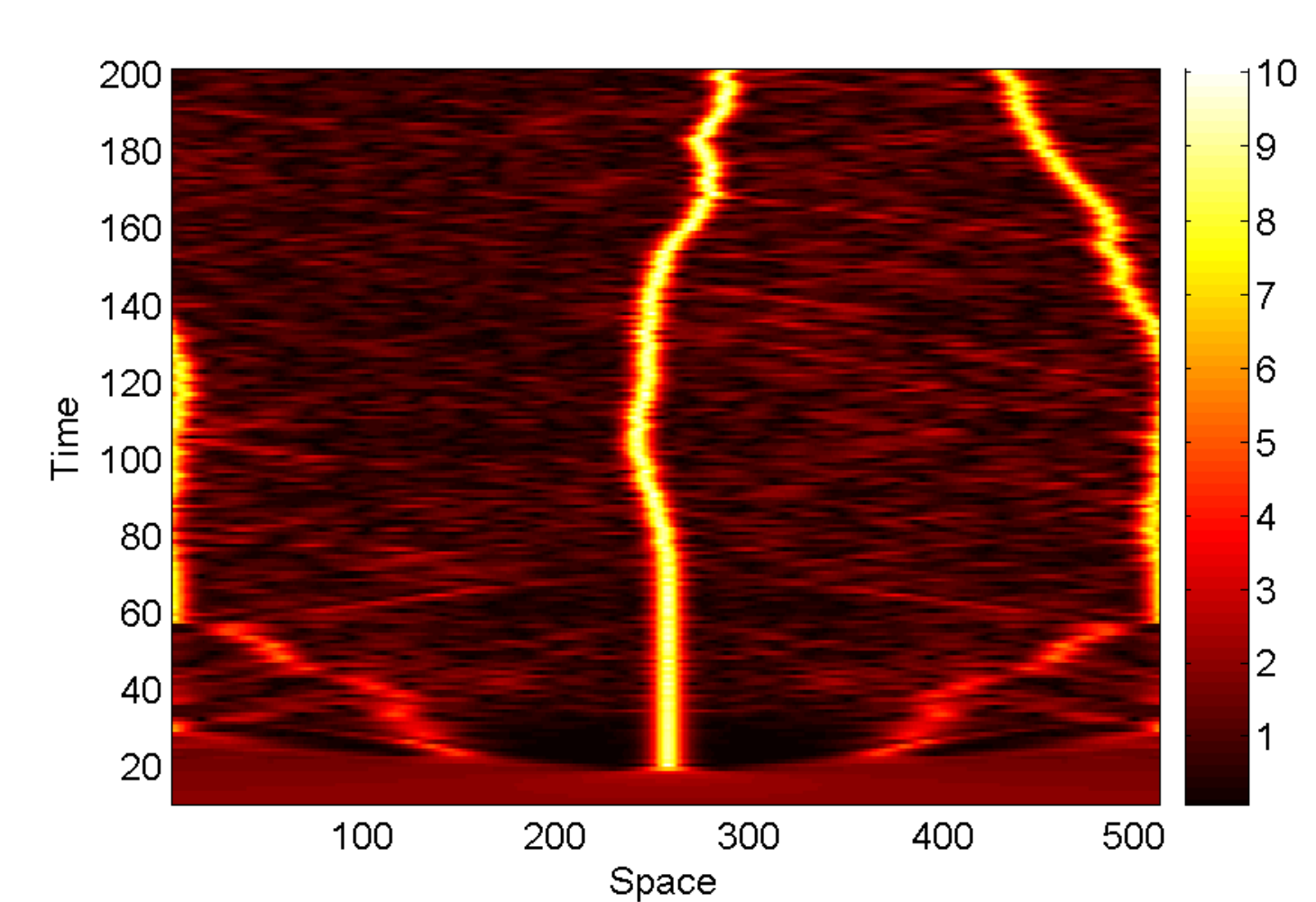}
\end{center}
\caption{(Color online) Contours of $|E(x,t)|=$const. with respect to the
space $(x\epsilon)$ and time $(t\epsilon)$ corresponding to the evolution as
in Fig. 7. This shows that the pattern selection leads to three harmonic
patterns. The system is in the coexistence of TC and SPC state.}%
\end{figure}\begin{figure}[ptb]
\begin{center}
\includegraphics[height=2.5in,width=2.5in]{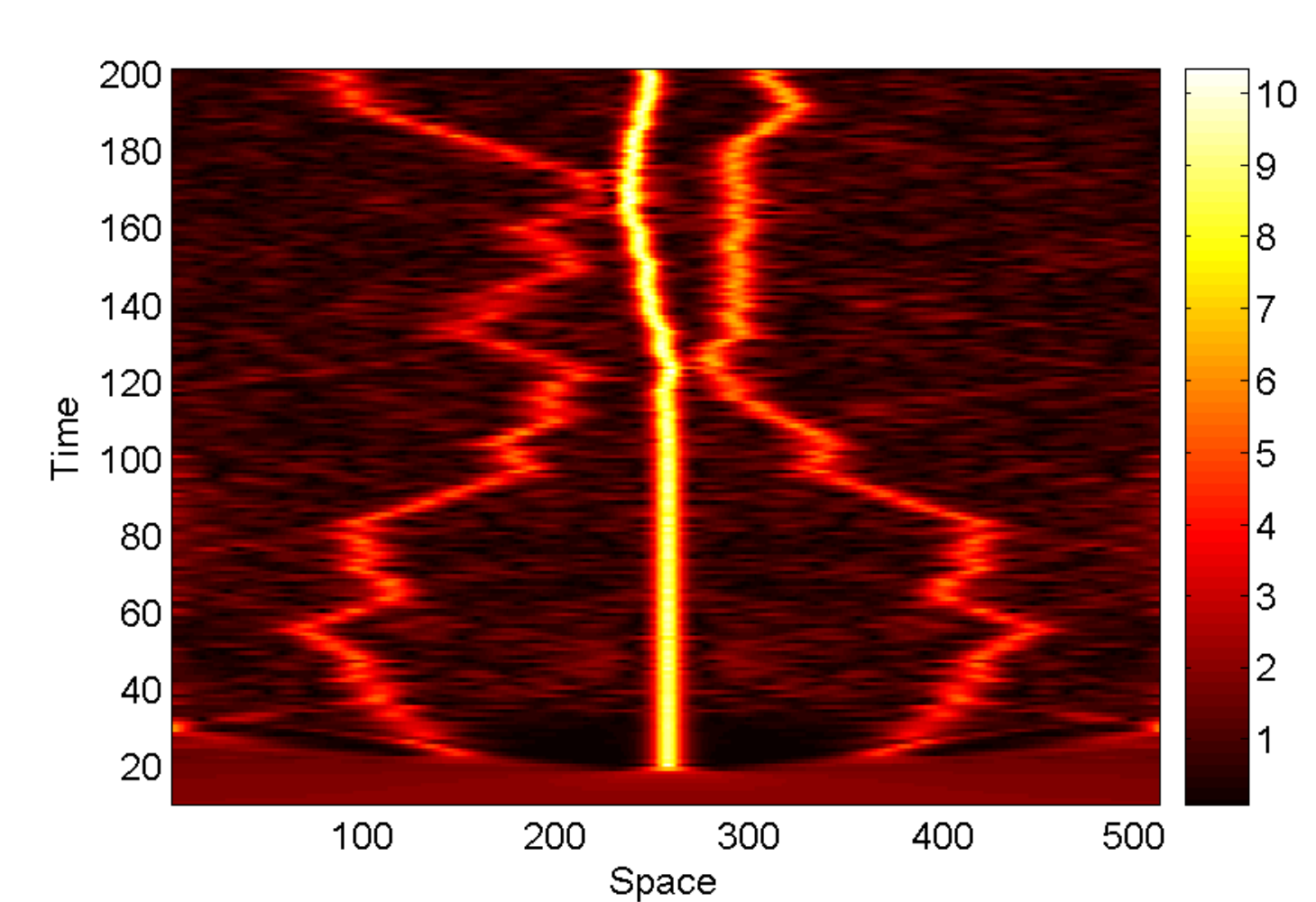}
\end{center}
\caption{(Color online) Contours of $|E(x,t)|=$const. with respect to the
space $(x\epsilon)$ and time $(t\epsilon)$ corresponding to the evolution as
in Fig. 8. This shows that the pattern selection leads to three harmonic
patterns. Two collisions occur at $t\approx120/\epsilon$ and $t\approx
170/\epsilon.$ The system is still in the coexistence of TC and SPC state.}%
\end{figure}

Now, if $k$ is further lowered, e.g., $k=0.065,$ many solitary patterns are
formed from the master mode and unstable harmonic modes (see Fig. 13 with IWN
and Fig. 14 without IWN) by means of pattern selection. We observe that when
IWN influences the dynamics, many unstable harmonic modes are excited compared
to the other case. As a result, not only two solitary patterns collide (as
mostly in the case of no IWN) to fuse into a new pattern, but more than two
also collide repeatedly to form new ones which again collide with other new
pattern(s) fused by the other collisions. Notice from Figure 13 that the first
collision occurs at $t\approx57/\epsilon$ between two solitary patterns. These
two patterns then fuse to form a new one, which again collides with other
three master patterns initially peaked at $x\approx190/\epsilon,$
$325/\epsilon$ and $700/\epsilon$ at $t\approx95/\epsilon,120/\epsilon$ and
$150/\epsilon$ respectively. There are also other collision and fusion that
can be explained similarly. In general, as time progresses, the patterns seem
to be much distorted in losing their strengths. The original $17$ (Fig. 13) or
$15$ (Fig. 14) solitary patterns are then finally fused into a few incoherent
ones. As solitons gradually vanish, irregular radiation increases due to
stochastic motion. That is, as soon as a chaotic subset with less degrees of
freedom is formed, the subset will then act as a pump delivering a net amount
of energy to its neighboring modes. This energy transfer takes place because
of the very random nature of the couplings. The system is then said to be in
the STC state. In this case, for $t>100/\epsilon$ the spatial correlation
function will approach to a zero value \ \cite{He}. A certain amount of
energy, which was initially distributed among the $15$ or $17$ solitary waves,
will now be transferred to a few incoherent patterns as well as\ to some
stable higher harmonic modes with short-wavelengths.

\ So, if initially there exist many modulational unstable modes with different
modulation lengths to form solitary envelopes, collision and fusion among most
of them can lead to the state of STC. There must exist a critical value of
this wavelength or wave number $k$ at which the transition from TC to STC
takes place. This value of $k$ is roughly in $0.03<k<0.1.$ Comparing the
pattern formation, collision and fusion in Figs. 11, 12 and 13, 14 one finds
that IAW emission is more stronger in presence of IWN. Thus, in order to
observe the interactions between LWs and IAWs at longer time scale and for
long-wavelength limits, the effect of IWN is more important, which can no
longer be neglected especially for relatively large amplitude waves. In this
way, the term containing the IWN in Eq. (\ref{e2}) would act as a correction
to the dynamics of arbitrary amplitude coupled Langmuir and IAWs.
\begin{figure}[ptb]
\begin{center}
\includegraphics[height=2.5in,width=2.5in]{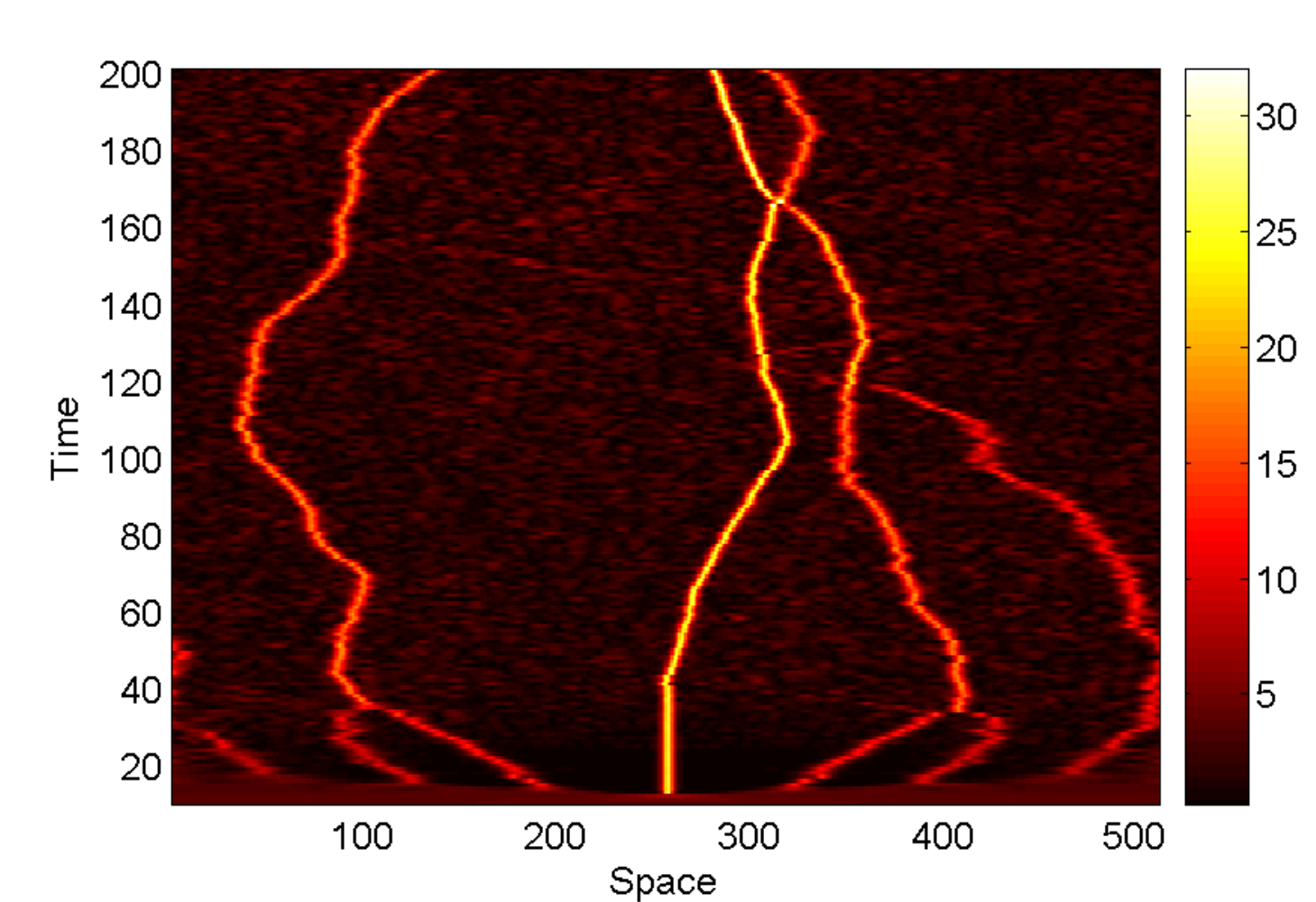}
\end{center}
\caption{(Color online) The same as in Fig. 9 for $k=0.22$, but for relatively
large $E_{0}$, i.e., $E_{0}=4$, indicating that more collision and fusion take
place due to IWN.}%
\end{figure}\begin{figure}[ptb]
\begin{center}
\includegraphics[height=2.5in,width=2.5in]{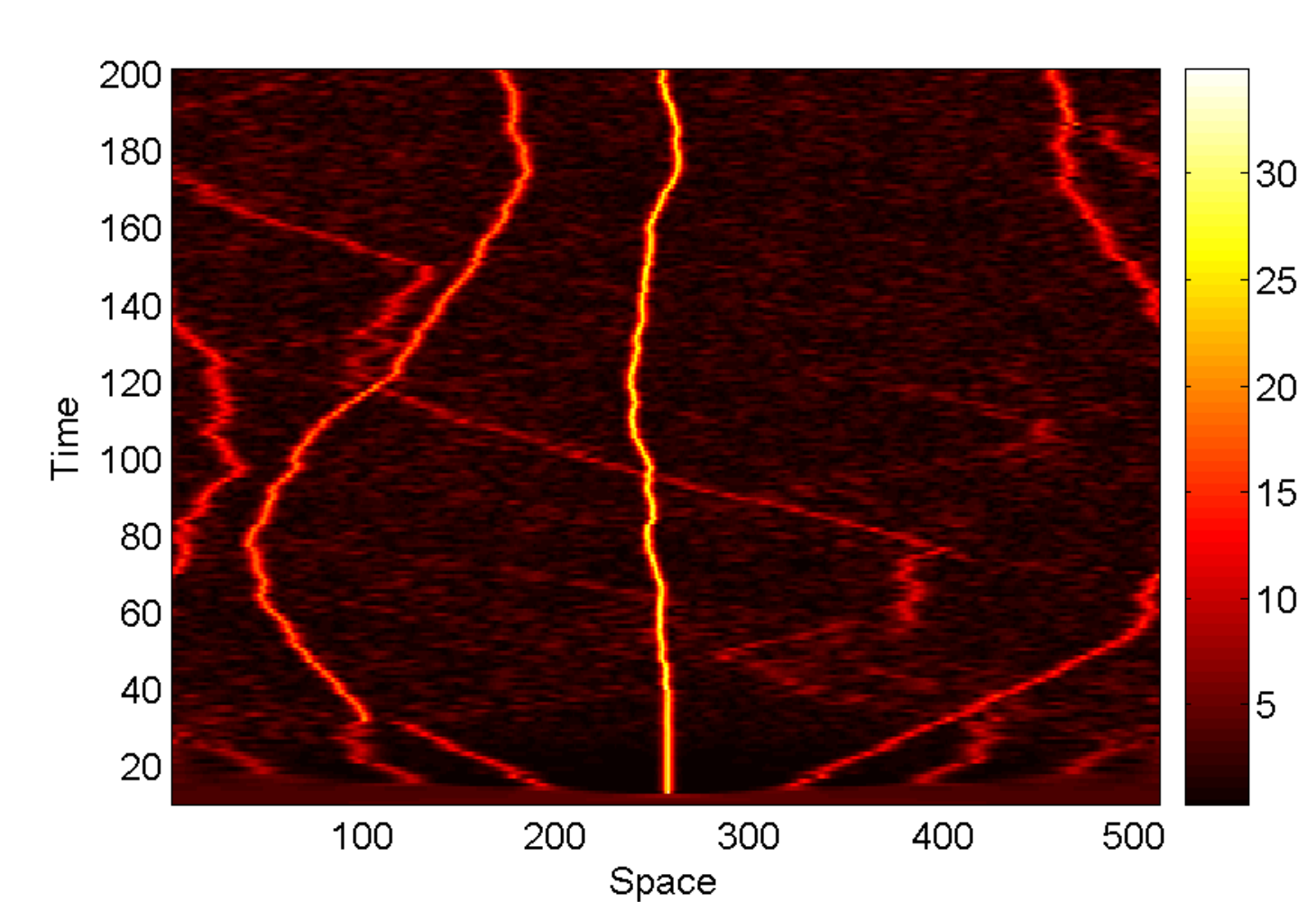}
\end{center}
\caption{(Color online) The same as in Fig. 11, but in absence of IWN,
indicating that relatively a less number of collision and fusion take place
compared to Fig. 11.}%
\end{figure}\begin{figure}[ptb]
\begin{center}
\includegraphics[height=2.5in,width=2.5in]{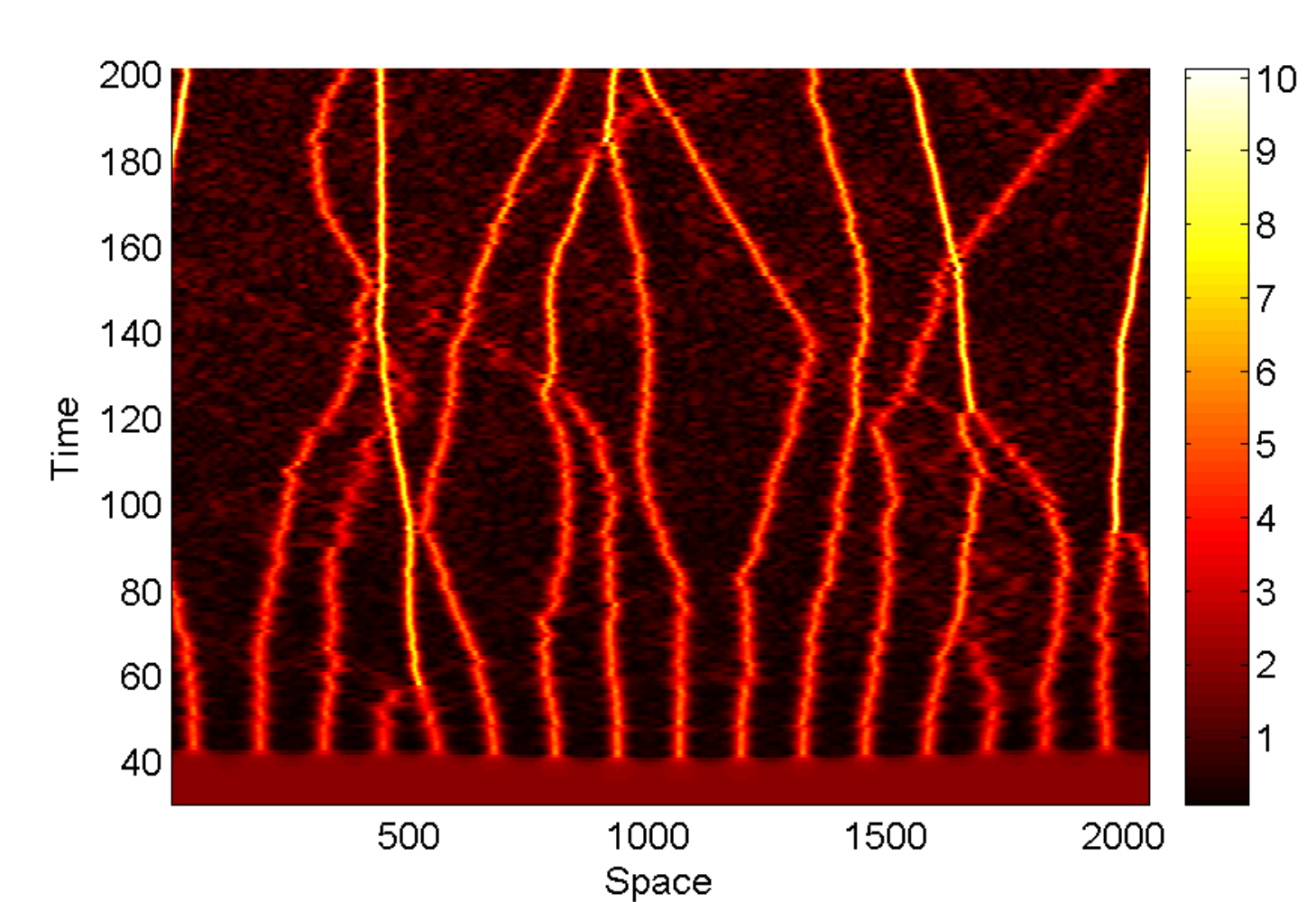}
\end{center}
\caption{(Color online) Contours of $|E(x,t)|=$const. with respect to the
space $(x\epsilon)$ and time $(t\epsilon)$ for $k=0.065$ and $E_{0}=2$ in
presence of IWN. The pattern selection shows that seventeen solitary patterns,
which were \ formed initially, collide and get fused into other incoherent
patterns after some time. The collision is random and not confined between two
patterns. The IAW emission also occurs, and the STC state emerges.}%
\end{figure}\begin{figure}[ptb]
\begin{center}
\includegraphics[height=2.5in,width=2.5in]{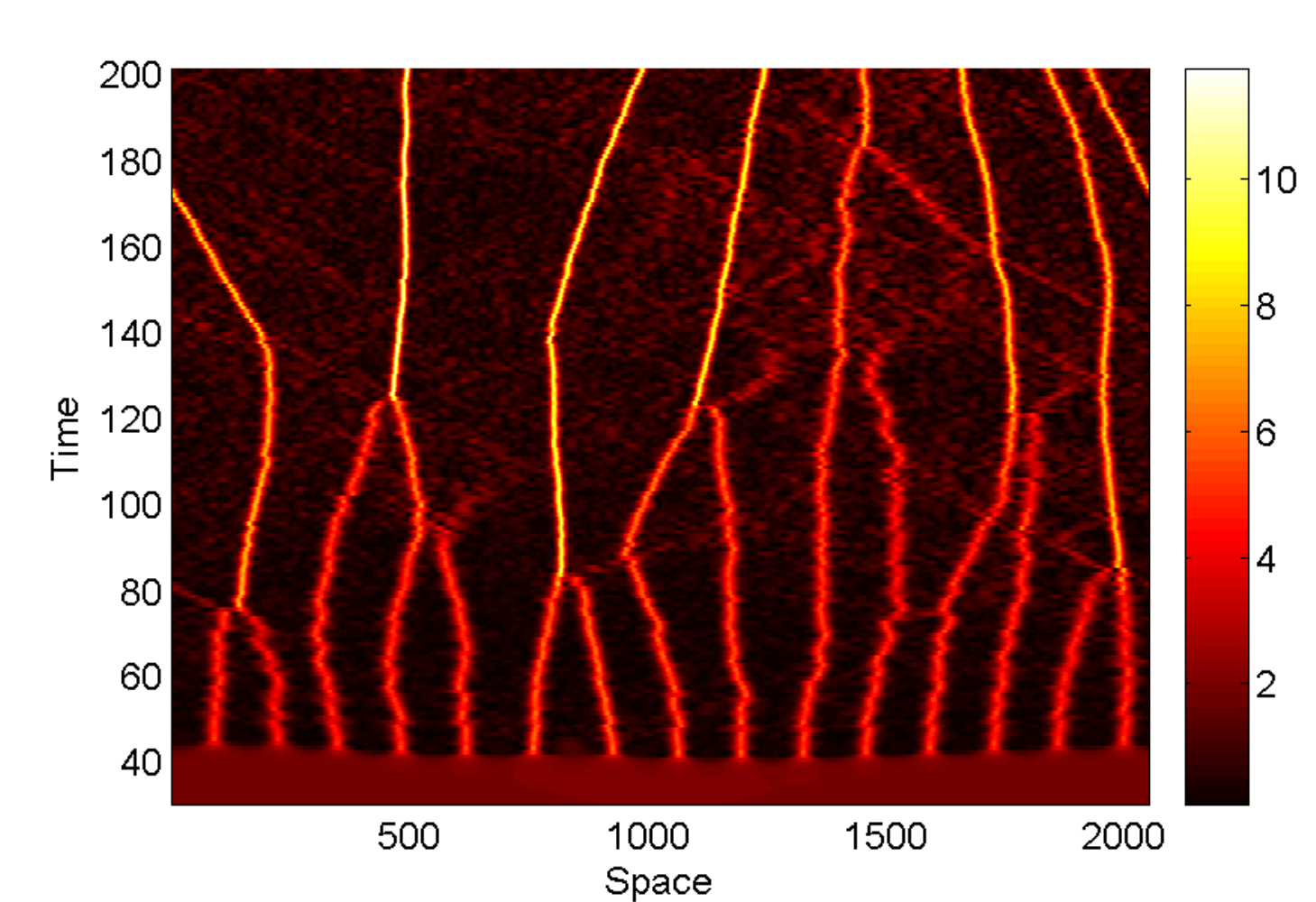}
\end{center}
\caption{(Color online) \ The same as in Fig. 13, but in absence of IWN.
Notice that fifteen solitary patterns were formed initially. After some time
they collide and fuse to form several new incoherent patterns. Here the
collision is mostly between two patterns compared to Fig. 13. The STC state is
also said to emerge in this case.}%
\end{figure}

\section{Discussion and conclusion}

From the analysis of the previous sections, we find that the three-wave model
can be a good approximation \cite{Rizzato} for the interaction of coupled LWs
and IAWs in the plane wave region $k_{c}/2<k<k_{c},$ where the system exhibits
stable oscillations. It can relatively be accurate in the region $0.3\lesssim
k<k_{c}/2$ where the system exhibits temporal chaos for a given value of the
pump electric field $E_{0}$. For values of $k<0.3,$ which indicates the
excitation of many ($>3$) unstable harmonic modes, the low-dimensional model
fails to describe the dynamics of the coupled waves. We have found that the
three-wave model exhibits periodic orbits in $1.2\lesssim k<k_{c},$and chaos
in $0.3\lesssim k\lesssim0.5<k_{c}/2$ for a given initial pump $E_{0}=2$ and a
given constant $N=6.$ The space-time evolution, on the other hand, reveals
that the system is in the state of TC for relatively a lower value of $k,$
i.e., in the regime $0.1\lesssim k\lesssim0.22,$ and the STC state emerges for
$0.03<k<0.1.$

To conclude, we have investigated the nonlinear interaction of coupled LWs and
IAWs by considering the effects of dispersion due to charge separation, and
the nonlinearity associated with the ion density fluctuations. The latter can
cause the excitation of more unstable harmonic modes, and can lead to strong
IAW emission in the space-time evolution. The low-dimensional model shows that
a transition from periodicity to chaos is possible for a suitable choice of
the wave number of modulation. It basically predicts some basic features of
the system as well as suggests an approximate region of $k$ for the existence
of regular and chaotic structures. The space-time evolution indicates that
\ solitary patterns are first formed by the unstable harmonic modes. These
modes are then excited by a master mode due to initial MI. \ As solitons
experience oscillatory motions, they emit radiation due to irregular
interaction with the ion-acoustic fields. Such radiations are broad-band in
nature, causing a growing number of modes to be involved in the stochastic
dynamics. Depending on the modulation length scale, the system may either be
in the TC and SPC states or in TC and STC states. If few harmonic patterns
coexist with the master pattern, and IAW emission remains weak, the system is
still in the state of TC and SPC. However, if many solitary patterns are
formed at the early stage due to MI, the system may experience \ the state of
SPC. Over longer time scales, when solitons are affected by the free
ion-acoustic radiation, collision and fusion among the patterns occur
frequently as well as repeatedly to form several new incoherent patterns. The
STC state is then said to emerge in which energy can flow faster the smaller
is the wave number of modulation $k$. Since our model is in one spatial
dimension, energy transfer is not a result of soliton collapse, rather it is
purely generated by the nonintegrable nature as well as chaotic aspects of the
system \cite{Rizzato}. The results may have physical significance to related
problems, such as turbulence in pulsar radiation where strong magnetic field
plays an important role \cite{Yu}.

S. B. is thankful to M.Pizzi of Micro and Nanotechnology Division, Techfab
s.r.l., Chivasso, Italy, for some useful discussion. A. P. M. gratefully
acknowledges support from the Kempe Foundations, Sweden.

\end{document}